\documentclass[journal]{IEEEtran}
\usepackage{graphicx}
\usepackage{cite}

\begin{document}

\title{Spatial mapping of the Dirac point in monolayer and bilayer graphene}
\author{Aparna~Deshpande,
        Wenzhong~Bao,
        Zeng~Zhao,
        Chun Ning~Lau,
        and~Brian~LeRoy%
 \thanks{Aparna Deshpande and Brian LeRoy are with the Department
of Physics, University of Arizona, Tucson,
AZ, 85721 USA e-mail: leroy@physics.arizona.edu}% <-this % stops a space
\thanks{Wenzhong Bao, Zeng Zhao and Chun Ning Lau are with the Department
of Physics, University of California, Riverside, Riverside, CA 92521 USA}% <-this % stops a space
\thanks{Manuscript received December 02, 2009}}

\maketitle

\begin{abstract}
We have mapped the Dirac point in exfoliated monolayer and bilayer graphene using spatially resolved scanning tunneling spectroscopy (STS) measurements at low temperature. The Dirac point shifts in energy at different locations in graphene. However, a cross correlation with the topography shows no correlation indicating that topographic features such as ripples are not the primary source of the variation.  Rather, we attribute the shift of the Dirac point to random charged impurities located near the graphene. Our findings emphasize the need to advance exfoliated graphene sample preparation to minimize the effect of impurities.
\end{abstract}

\begin{IEEEkeywords}
graphene, dirac point, STM, spectroscopy
\end{IEEEkeywords}

\section{Introduction}
Graphene is the two dimensional form of carbon with novel structural, optical and electrical properties that make it a promising material for device applications\cite{geim2009,neto2009}. Graphene can be accessed as a single sheet of carbon atoms, monolayer graphene(MLG) or as two stacked sheets of carbon atoms, bilayer graphene(BLG).  Both types of graphene are gapless semiconductors. MLG has a linear band structure and the charge carriers can have very high mobility exceeding 200,00 cm$^2$/Vs\cite{bolotin}.  This high mobility which greatly exceeds the values found in Si opens the way to new type of electronic devices from MLG.  However, these high values of mobility are only found in suspended samples implying that supported samples have sources of scattering which limits their mobility.  BLG, on the other hand, has a parabolic band structure and a quadratic energy dispersion. Because of the reduced symmetry of the BLG, applying an electric field perpendicular to the plane of BLG opens a band gap. This band gap tunability makes BLG a very promising material for digital electronics applications\cite{zhang2009BLG}.

The isolation of MLG and BLG from graphite, the three dimensional form of carbon, gave rise to the field of graphene experiments and theory. The exfoliated sheets are then studied by depositing them on a silicon dioxide substrate. Graphene can also be prepared by epitaxy on silicon carbide substrates\cite{berger2004}, irridium\cite{coraux2008} and ruthenium\cite{sutter2008} as well as chemical vapor deposition growth on copper\cite{li2009} and nickel\cite{kim2009}. Every technique has its own advantages and disadvantages. Epitaxial graphene needs to be delocalized enough from the substrate to be useful for nanoelectronic devices.   However, these growth techniques tend to provide much larger samples than the exfoliation technique, up to wafer size in the case of silicon carbide which is useful for commercial applications.  Current growth techniques produce large samples which are not uniform in thickness and show reduced quality compared to exfoliated samples.  Exfoliated graphene has been shown to have intrinsic ripples and not be atomically flat \cite{meyer2007,fasolino2007}.  These ripples can influence the electronic properties by changing the nearest neighbor hopping amplitude.  This variation in the hopping amplitude leads to a shift in the electrochemical potential and hence the Dirac point \cite{kim2008}.  Exfoliating graphene on silicon dioxide substrates leads to random charged impurities on the sheet due to impurities trapped between the oxide layer and graphene sheet. These impurities also influence the electrical properties of graphene and may be the source of the reduced mobility in supported samples. To examine the effect of ripples and charged impurities on exfoliated MLG and BLG we have performed measurements using a low temperature scanning tunneling microscope (STM).

\section{Sample preparation}
Mechanical exfoliation of graphene, a seemingly simple but intricate technique \cite{novoselov2004,novoselov2005} was used to prepare our samples. Graphene flakes were then transferred to the SiO$_2$ substrate. Ti/Au electrodes were deposited on the graphene using a shadow mask technique described elsewhere \cite{LauSub}.  The use of a shadow mask for the electrodes eliminates the need to use PMMA as a resist which is a possible source of charged impurities.  Eliminating the PMMA has also been seen to increase the cleanliness of the graphene devices.  This is evidenced by atomic resolution topography of freshly prepared samples without the need for high temperature cleaning.  The cleanliness is also seen in an improved mobility in samples fabricated with the shadow mask technique.  The samples were transferred to the ultrahigh vacuum (p $\leq 10^{-11}$ mbar) STM and cooled to 4.6 K. Electrochemically etched tungsten tips were used for imaging and spectroscopy. These tips were checked on a Au surface to ensure a constant density of states prior to measurements on graphene.

\section{Measurements}
Figure 1(a) shows the setup for scanning tunneling spectroscopy measurements on graphene flakes. The voltage is applied to the tip attached to the piezo tube and the sample is grounded. STM images show the hexagonal lattice characteristic of MLG \cite{deshpande2009,zhang2008,zhang2009} and the triangular lattice for BLG \cite{stolyarova2007,ishigami2007,deshpande2} as shown in Fig. 1(c) and 1(b) respectively.  BLG shows a triangular lattice because the two layers of carbon atoms reduce the symmetry of the graphene and therefore only one of the two sublattices is visible.

We have recorded the local density of states for both MLG and BLG with dI/dV point spectroscopy measurements. Here the tip is fixed at a definite position on the sample, the feedback loop is deactivated and the sample voltage is ramped within an energy window. An ac modulation voltage of 5 mV rms, 570 Hz is applied to the tip and the resulting dI/dV spectrum is recorded using lockin detection. Fig. 1(d) shows one such dI/dV spectrum for BLG while the results for MLG are in Fig. 1(e). The minimum of each spectrum corresponds to the local Dirac point.  In the BLG, this minimum clearly occurs away from the Fermi energy which is at 0.0 V.  This is evidence that the local electrochemical potential has shifted but the origin of this shift is not clear without further analysis as discussed below.

\begin{figure}[t]
\centering
\includegraphics[width= 0.45 \textwidth]{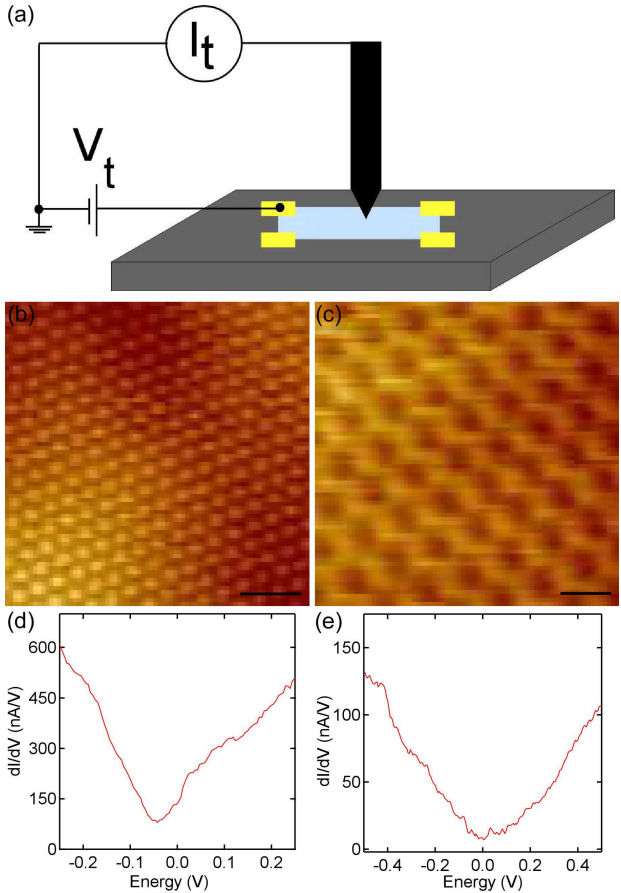}
\caption{\label{fig:schematic} (color online) (a) Measurement setup showing the graphene flake (in blue) on a SiO$_2$ substrate along with the STM tip and its electrical connections. (b) An image of bilayer graphene showing the triangular lattice, (0.5 V, 100 pA). The scale bar is 7{\AA}. (c) An image of monolayer graphene with the characteristic hexagons, the scale bar is 3{\AA}, (0.5 V, 100 pA). (d) dI/dV spectroscopy for bilayer graphene showing the Dirac point (0.25 V, 100 pA). (e) dI/dV spectroscopy for monolayer graphene showing the Dirac point. (0.5 V, 50 pA)}
\end{figure}

Previous STM studies have shown that in case of MLG impurities \cite{deshpande2009,zhang2009} and phonons \cite{zhang2008} influence the charge-carrier scattering mechanisms in graphene.  There has also been the observation of electron and hole puddles due to localized charged regions of the MLG.\cite{deshpande2009,zhang2009,martin2008}.  However, the connection between the localized charge puddles and the topography of the graphene has not been firmly established.

Here we have measured the spatial variation of the Dirac point in order to correlate it with the local topography. Figure 2(a) shows the topography recorded for a 100 nm $\times$ 100 nm area of MLG. For this area, spectroscopy measurements were carried out on a 1 nm $\times$ 1 nm grid with an energy resolution of 5 meV.  These measurements were used to create a local density of states (LDOS) map. From this map, we have calculated the shift of the Dirac point at each point based on the minimum in the density of states.  The energy shift of the Dirac point as a function of position is plotted in Fig. 2(b). The bright regions correspond to a positive shift and the dark regions represent a negative shift. The average shift is 88 mV and the standard deviation of the shift is 84 mV. The average shift indicates that the Dirac point is shifted away from the Fermi energy while the variation indicates the strength of the puddles.  The spatial extent of the puddles is approximately 8-15 nm in size. These sizes match well with the roughness of the SiO$_2$ substrate \cite{stolyarova2007,ishigami2007,geringer2009}.

\begin{figure*}[t]
\centering
\includegraphics[width=1.0 \textwidth]{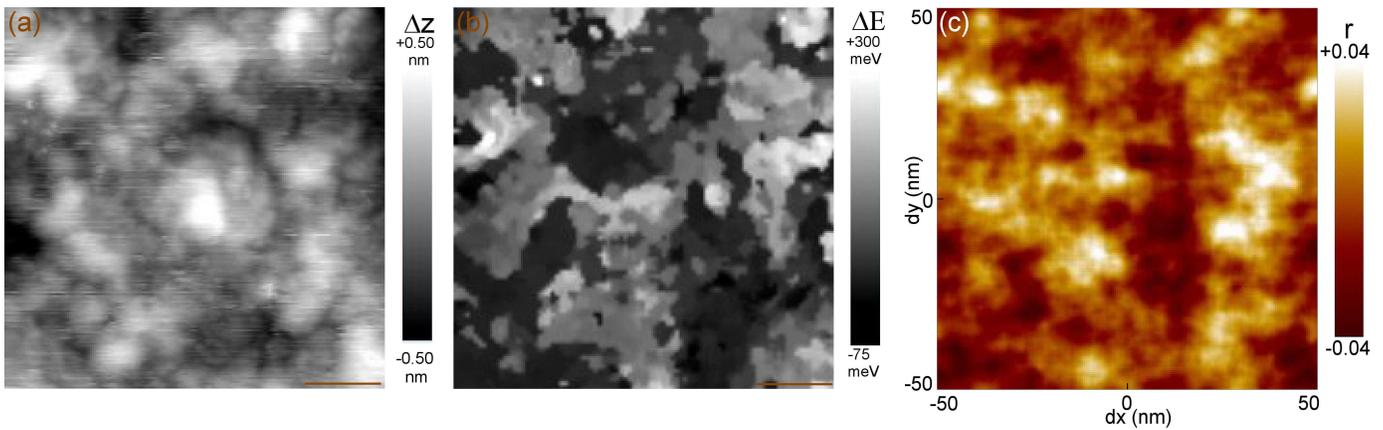}
\caption{\label{fig:2Dpoint} (color online) (a) Topography of monolayer graphene (0.5 V, 50 pA). The scale bar is 20 nm. (b) Image showing the shift of the Dirac point calculated from dI/dV spectroscopy map recorded in the same region as the topography shown in (a). The scale bar is 20 nm. (c) Cross correlation between the curvature calculated from the topography in (a) and the Dirac point shift in (b). The color scale bar corresponds to the cross correlation coefficient r.}
\end{figure*}

Theory calculations have predicted that the ripples in graphene generate an electrochemical potential which varies as the square of the local curvature of the graphene flake \cite{kim2008}. To investigate the influence of ripples in our measurements, we used the topography image (Fig. 2(a)) to calculate the curvature of the graphene and thereby get a map of the electrochemical potential generated by the ripples. Then we carried out a cross correlation of this electrochemical potential with the map of the Dirac point shift (Fig. 2(b)). The cross correlation image is shown in Fig. 2(c). If the shift in our measurements were to be a consequence of the ripples, it would be manifested as a peak in the center of the cross correlation image. However there is no peak or any structure in the cross correlation image (Fig. 2(c)). In all locations the degree of correlation, r$^2$ is less than 0.5$\%$.  This indicates that there is no correlation between the topography and the Dirac point shift in the case of MLG.

We have repeated the analysis for BLG.  Figure 3(a) shows the topography for a 60 nm $\times$ 60 nm area of BLG. Spectroscopy measurements were taken on a 0.6 nm $\times$ 0.6 nm grid with a 5 meV energy resolution. Figure 3(b) shows the plot of the shift of the Dirac point. Here the average shift is -44 mV and the standard deviation of the shift is  27 mV. The size scale of the puddles is once again approximately 8-15 nm. Similar to MLG, we have calculated the curvature and electrochemical potential variation from the topography image. This image is then cross correlated with the Dirac point shift image to produce Fig. 3(c). Once again the two images show no correlation as the correlation coefficient for the unshifted images is approximately the same as for any shift of the two images.

\begin{figure*}[t]
\centering
\includegraphics[width=1.0 \textwidth]{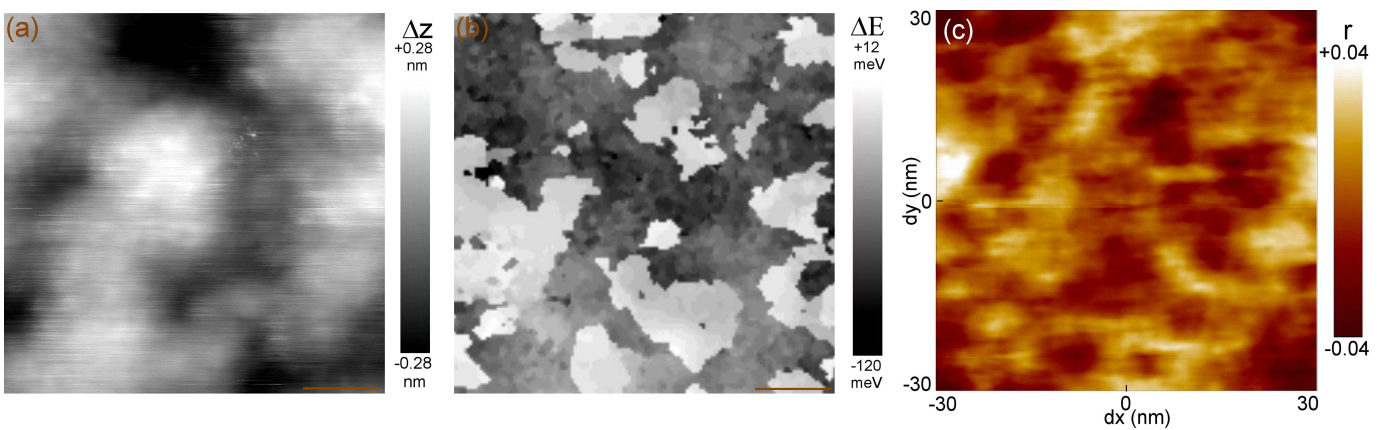}
\caption{\label{fig:map} (color online) (a) Topography of bilayer graphene (0.25 V, 100 pA). The scale bar is 10 nm. (b) Image showing the shift of the Dirac point calculated from a dI/dV spectroscopy map recorded in the same region as the topography shown in (a). The scale bar is 10 nm. (c) Cross correlation between the curvature calculated from the topography in (a) and the Dirac point shift in (b). The color scale bar corresponds to the cross correlation coefficient r.}
\end{figure*}

\section{Conclusion}
We have mapped the Dirac point by spatially resolved spectroscopy for monolayer and bilayer graphene using low temperature scanning tunneling microscopy. Using cross correlation we have shown that there is no correlation between the topography and spectroscopy images for monolayer and bilayer graphene. Therefore, we can rule out curvature as the mechanism for creating the shift in the electrochemical potential.  Hence, we attribute the shift of the Dirac point to random charged impurities. A recent technique of making graphene on mica substrates instead of silicon dioxide seems promising in terms of reducing the ripples \cite{Lui2009}. Spectroscopy measurements need to be carried out with exfoliated graphene on mica substrates to check the effect of impurities.  Further studies can also be done on suspended graphene devices where the amount of ripples increases but the oxide substrate is no longer present.  More efforts are necessary in this direction to develop an ideal substrate for using graphene in nanoelectronic devices.

\section*{Acknowledgment}
AD and BJL acknowledge the support of the U. S. Army Research Laboratory and the U. S. Army Research Office under contract/grant number W911NF-09-1-0333 and National Science Foundation grant ECCS/0925152.  WB, ZZ and CNL acknowledge support by NSF CAREER DMR/0748910, NSF/ECCS 0926056, ONR N00014-09-1-0724 and ONR/DMEA H94003-09-2-0901.

\bibliographystyle{IEEEtran}

\end{document}